\documentclass[aps,prb,floatfix,twocolumn,showpacs]{revtex4}
\usepackage{amsmath,amssymb,graphicx,bm,epsfig,color}

\begin{document}

\title{X-ray absorption branching ratio in actinides: LDA+DMFT approach}

\author{J. H. Shim, K. Haule, and G. Kotliar}

\affiliation{Department of Physics, Rutgers University, Piscataway,
             NJ 08854, USA}

\begin{abstract}
To investigate the x-ray absorption (XAS) branching ratio from the
core $4d$ to valence $5f$ states, we set up a theoretical framework
by using a combination of density functional theory in
the local density approximation and Dynamical Mean Field Theory
(LDA+DMFT), and apply it to several actinides.
The results of the  LDA+DMFT reduces to the
band limit for itinerant systems  and  to the atomic limit for
localized $f$ electrons, meaning a spectrum of $5f$ itinerancy
can be investigated. 
Our results provides a consistent and unified view
of the XAS branching ratio for all elemental actinides,  and is in
good overall agreement with experiments.
\end{abstract}

\date{\today}
\maketitle

%{\bf General  Introduction}
Understanding the physics of  elemental actinide solids  is
an important issue for many body physics as well as for applications in
nuclear power generation.
In the early actinides, the $f$ electrons behave as waves
delocalized through the crystal,
while in the late actinides, the $f$ electrons behave as particles
localized around each atom.
Plutonium is near the localization-delocalization edge separating
these two regimes.
X-ray absortion (XAS)  from the core $4d$ to the
valence $5f$ in conjunction with atomic physics calculations has
been a powerful probe of the evolution of the  valence and the
strength of the spin-orbit coupling across the actinide series.

%WHY PU IS INTERSETING.
% The study of actinide material has been
%important issue for long time because of the strong correlation
%effect of $5f$ electrons, as well as its industrial interest for
%nclear reactors. Due to its localized distribution of orbital and
%strong spin-orbit interaction, the $5f$ electrons show various
%anomalous behavior, even in the elemental metals. As increasing
%atomic number of elemental metal, they show delocalized and
%localized transition of $5f$ electrons, which induces diverse
%anomalous properties in transport, magnetism, specific heat, and
%structures. To understand this transition of actinides, it is
%important to describe the competition of spin-orbit interaction
%and other effect.

%ACTINDES HAVE BEEN STUDIED WITH DIFFERENT SPECTROSCOPIES. THEY
%PROBE DIFFERENT ASPECTS OF THE VALENCE BAND. DIFFERENT ENERGY
%PROBES ACCESS THE WAVE OR PARCTICLE ASPECT OF THE ELECTRON.

%A LOT OF USEFUL INFO HAS BEEN GAINED BY COMBINING BRANCHING RATIO
%AND ATOMIC MULTIPLET THEORY: APPROXIMATE f COUNT, STRENGTH

A large number of spectroscopies have been applied to this
problem. For example, photoemission spectroscopy of Pu, has
revealed  a multiple-peak structure in the occupied part of the
density of states\cite{Arko00,Tobin03,Gouder05}.
A combination approach of local density approximation (LDA) and Dynamical
Mean Field Theory (LDA+DMFT) has allowed the interpretation of
these  features in terms of $f$ electrons which are delocalized at
low frequencies with  a mixed valent electron count\cite{Shim}.
Other interpretations of Pu spectroscopies using LDA+DMFT with other
impurity solvers have been presented recently\cite{Shick07,Zhu07,ctqmc}.

High-energy probes such as electron energy-loss spectroscopy (EELS) or x-ray
absorption spectroscopy (XAS) constitute a different set of
spectroscopies which have been intensively used to study the
electronic structure of the actinide
series\cite{vanderLaan96,vanderLaan04,Tobin05,Moore06,Moore07,Moore07b,Butterfield08}.
These experimental works, combined with
theoretical calculations exploiting a powerful sum rule, and the
electronic structure of the atom has yielded  valuable insights on
the degree of localization of valence $5f$ electrons in
actinides\cite{vanderLaan04,Tobin05,Moore07b} as well as
the spin-orbit strength\cite{vanderLaan04,Moore07,Moore07b}.

\begin{figure}[tbh]
\includegraphics[width=0.8\linewidth]{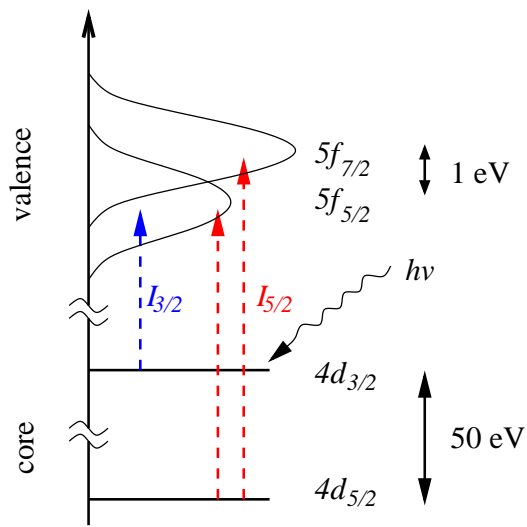}
\caption{
Schematic diagram of the XAS from core $4d$ to valence $5f$ transition.
The absorption intensity $I_{5/2}$ and $I_{3/2}$ correspond
to $4d_{5/2} \rightarrow 5f_{5/2,7/2}$ and $4d_{3/2} \rightarrow 5f_{5/2}$
transitions, respectively.
The core levels are clearly discretized by $4d_{3/2}$ and $4d_{5/2}$ level,
but the valence levels are overlaped between $5f_{5/2}$ and $5f_{7/2}$ level
due to mltiplet splittings and electron itinerancy.
%If the spin-orbit interaction is much smaller than other effect such as
%Hund's rule coupling, crystal field effect, band broadening, the branching ratio
%is given by a statistical value of 3/5.
%If the spin-orbit interaction is comparable to
%other effects, the branching ratio can be correctly treated
%by atomic model considering spin-orbit interaction and Hund's coupling.
}
\label{xas}
\end{figure}

Here we address the computation of the branching ratio within
DMFT. There are several motivations for this  study. 1)   While
the atomic multiplet approach of Ref.~\onlinecite{vanderLaan96}
describes very successfully the majority of the data in the late
actinides, it is restricted to the  case that the $f$ electrons are
strictly localized.
% and therefore can be treated using  atomic
%multiplet theory.
It is  therefore useful to extend  this approach
by embedding it  in a more general method that captures the
itinerant limit as well. 2) Many other spectroscopies of the
actinides such as photoemission spectroscopies   are not well
described by a localized model. LDA+DMFT provides a unified
picture which reconciles the results of high-energy and low-energy
spectroscopies. In physical terms, DMFT can  describe the low-energy
part of the excitations as itinerant, and the high-energy
excitations as localized. Different spectroscopies weight
different parts of the spectra.  It is useful to
incorporate the XAS branching ratio in the general LDA+DMFT
formalism to achieve a unified interpretation of high energy and
low energy spectroscopies.  3) The branching ratio can be
expressed as a ratio of two quantities, one involving the spin-orbit
coupling and the other involving the $f$ occupation.
Therefore a proper theoretical interpretation of the experiments
requires the evaluation of these two quantities. 
Until now, experimentally the occupancies were determined to be
in an interval of allowed values. 
%Until now,
%occupancies were assumed to be either integers or
%some range of values. 
XAS and EELS experiments
were  used to constrain the variations of the spin-orbit coupling
for that range of occupancies.

%Recent experiments of XAS and EELS on actinide elements have
%shown the full database of branching ratio value, which enable
%the systematic study of the spin-orbit interaction. However, in
%experiments, the number of hole $\langle n_h \rangle$ is not
%exactly known, and only the ratio between spin-orbit strength and
%$\langle n_h \rangle$ is known from the sum rule. So, an
%complementary work has been needed from theoretical approach.

%{\bf Introduction to Branching Ratio}
Figure~\ref{xas} shows a schematic diagram of the $4d \rightarrow
5f$ transition in  XAS. K. Moore and collaborators have shown that EELS
experiments give equivalent
information\cite{rmp_Moore,Moore04}. One envisions a large splitting due to
spin-orbit coupling of the core levels. The  electric-dipole selection
rule ($\Delta j = 0, \pm 1$), gives rise to two  absortion lines
%kinds of transition intensity
: $I_{5/2}$ ($4d_{5/2} \rightarrow
5f_{5/2,7/2}$) and $I_{3/2}$ ($4d_{3/2} \rightarrow 5f_{5/2}$).
The branching ratio $B$ is defined by
$B=I_{5/2}/(I_{5/2}+I_{3/2})$.

When the spin-orbit splitting in the valence band 
is negligible in comparison with the
different sources of  broadening (due to multiplet splittings or
electron itineracy)  shown schematically in Fig.~\ref{xas}, we
can neglect the spin-orbit splitting in the final states and the
branching ratio $B$ is given by a statistical value of 3/5, which
reflects the relative degeneracies of the initial states. Hence
the branching ratio is a probe of the strength of the spin-orbit
coupling\cite{vanderLaan04,Moore07,Moore07b}.

%In localized system, on the other hand, it can be described by
%atomic model considering the spin-orbit interaction and Hund's
%coupling correctly. As a result, the measurement of branching
%ratio enable us to probe the localization of $5f$ electrons.
%

%When the spin-orbit interaction is much larger than other effects,
%such as Hund's coupling, crystal field effect, and Kondo screening,
%they are called $jj$ coupling. On the other hand, if the spin-orbit
%interaction is minor in the valence states, the branching ratio
%is dominated by statistical value, and they are called $LS$ coupling.
%In isolated atom, their behavior is in between theses $LS$ and $jj$
%coupling scheme, which are called intermediated coupling scheme.

Detail calculations based on the isolated atomic model were
shown to be  consistent with the experimental result of the
late actinides. However,  deviations get larger when the
system becomes delocalized as in U and Np. Due to the
delocalization, the spin-orbit strength is apparently suppressed.
We will show that  although LDA  is a proper approximation for
describing many properties of
 itinerant system, it  underestimates the experimental
results of early actinides. So more realistic theoretical model,
which can treat both the atomic and itinerant physics, is needed.
The LDA+DMFT method is the most promising method for this purpose.

%{\bf Derivation}
The spin-orbit sum-rule has been fully derived  in  the pioneering
work of van der Laan, Thole and collaborators
\cite{vanderLaan96,thole88b,vanderLaan88,thole88a},
and applied to numerous systems. (for recent review,
see Ref.~\onlinecite{rmp_Moore}.) Here we summarize the main
steps of their derivation, indicating the  places where the atoms
are considered in a solid state environment, and where a treatment
going beyond  band theory and atomic multiplet physics, such as
Dynamical Mean Field Theory (DMFT) is required.

The X-ray  absorption resulting from  a core-valence transition,
is described by  a term in the Hamiltonian that couples the
electromagnetic field to a transition operator $T_q$.
% on top of
%the LDA+DMFT hamiltonian. Using
which in the electric dipole approximation is given by:
\begin{eqnarray}
T_q = \sum_i T_q(i),
\end{eqnarray}

$T_q(i)$ is the atomic operator at site $i$. We consider
transition  from a core level denoted by $j_c$ $(= l_c \pm s)$ to
the partially occupied  valence levels denoted by $j_v$ $(= l_v
\pm s)$  with  absorption of  $q$ polarized light.
%is generalized in the lattice problem.
%
%

\begin{eqnarray}
T_q(i) = \sum_{m_v m_c}\langle l_v s;j_v m_v| \mathbf{r}_{q}|l_c s;j_c m_c \rangle
f^{\dagger}_{j_vm_v} d_{j_cm_c},
\end{eqnarray}
where $d_{j_c m_c}$ is the annihilation operator of a core electron
and $f^{\dagger}_{j_v m_v}$ is the creation operator of a valence electron.

The matrix element can be given as a product of a reduced matrix
element containing  a radial integral and  an angular dependent
part\cite{vanderLaan98}.
\begin{multline}
\langle l_v s ; j_v m_v|\mathbf{r}_{q}|l_c s ; j_c m_c \rangle
  =(-1)^{j_v-j_c} [l_v l_c j_c]^{1/2}\\
\times
\left\{ \begin{array}{c c c}
j_c & 1 & j_v \\
l_v & s & l_c
\end{array} \right\}
\left( \begin{array}{c c c}
j_c & 1 & j_v \\
m_c & {q} & -m_v
\end{array} \right)
\langle l_v s || \mathbf{r} || l_c s \rangle,
\label{wigner}
\end{multline}
where $[a,b,\cdots]$ is shorthand for $(2a+1)(2b+1)\cdots$.
Here, we used the Wigner-Eckart theorem. Because
we are interested in the transition of given core ($l_c$) and valence ($l_v$)
states, we will omit the reduce matrix element below.

When we consider the isotropic spectrum, where the light polarizations
are averaged,
the absorption intensity summed over the final states $|f\rangle$
from a many-electron ground state $|g\rangle$ is
\begin{multline}
I = \sum_q \sum_f \langle g |T^{\dagger}_q| f\rangle
                  \langle f | T_q         | g \rangle \\
  = \sum_q \sum_{m_v m_v'} \sum_{m_c m_c'}
    \langle  d^{\dagger}_{j_c m_c'}f_{j_v m_v'}
    f^{\dagger}_{j_v m_v}d_{j_c m_c} \rangle \\
 \times \langle l_c s ; j_c m_c'|\mathbf{r}_{q}|l_v s ; j_v m_v' \rangle
        \langle l_v s ; j_v m_v|\mathbf{r}_{q}|l_c s ; j_c m_c \rangle.
\end{multline}
Here we assume that the interference terms  ($\langle T_q(i)
T_q(j)\rangle$, $i\neq j$) are  negligible which is a valid
assumption when the core electron states are effectively
localized. Since $|g\rangle$ does not contain holes in the core
level, the core shell operators are removed by $d^{\dagger}_{j_c
m_c'}|g\rangle=0$, and the creation-annihilation term in the
intensity can be obtained as
\begin{equation}
    \langle d^{\dagger}_{j_c m_c'}f_{j_v m_v'}
    f^{\dagger}_{j_v m_v}d_{j_c m_c}\rangle
    = \langle f_{j_v m_v'}f^{\dagger}_{j_v m_v} \rangle
      \delta_{m_c m_c'}
\end{equation}
Where we  neglected the core valence hybridization and  we neglect
the core-hole valence interaction, which is a reasonable
approximation in $4d$ $\rightarrow$ $5f$ transition.
\cite{vanderLaan04}

Combining Eqs. (3)-(5),
the the dipole transition probability is
\begin{multline}
I = [l_v l_c j_c]
    \left\{ \begin{array}{c c c}
    j_c & 1 & j_v \\
    l_v & s & l_c
    \end{array} \right\}^2
    \sum_{m_v m_v'}
    \langle f_{j_v m_v'}f^{\dagger}_{j_v m_v} \rangle \\
    \times
    \sum_{m_c q}
    \left( \begin{array}{c c c}
    j_c & 1 & j_v \\
    m_c & {q} & -m_v
    \end{array} \right)
    \left( \begin{array}{c c c}
    j_c & 1 & j_v' \\
    m_c & {q} & -m_v'
    \end{array} \right) \\
    = [l_v l_c j_c]
    \left\{ \begin{array}{c c c}
    j_c & 1 & j_v \\
    l_v & s & l_c
    \end{array} \right\}^2
    \langle n^h_{j_l} \rangle.
\end{multline}

We have used the normalization condition.
\begin{multline}
    \sum_{ m_c q}
    \left( \begin{array}{c c c}
    j_c & 1 & j_v \\
    m_c & {q} & -m_v
    \end{array} \right)
    \left( \begin{array}{c c c}
    j_c & 1 & j_v' \\
    m_c & {q} & -m_v'
    \end{array} \right) \\
    = [j_v]^{-1} \delta_{j_v j_v'} \delta_{m_v m_v'}
\end{multline}

We consider the dipole transition from core $4d$ states ($j_c = 2\pm 1/2$)
to valence $5f$ states ($j_v = 3 \pm 1/2$).
Using the relevant values of 6-$j$ symbol in Eq. (5),
the absorption intensity for each transition is given as
\begin{eqnarray}
I_{3/2} =
I(4d_{3/2} \rightarrow 5f_{5/2}) =  \langle n^h_{5/2} \rangle
(2l_v+1)(l_v-1)/l,
\end{eqnarray}
\begin{multline}
I_{5/2} =
I(4d_{5/2} \rightarrow 5f_{5/2})
+I(4d_{5/2} \rightarrow 5f_{7/2}) \\
=  \langle n^h_{5/2} \rangle /l_v
+  \langle n^h_{7/2} \rangle (2l_v-1).
\end{multline}
The branching ratio for $4d \rightarrow 5f$ transition
is given as
\begin{multline}
B = \frac{I_{5/2}}
      {I_{5/2} + I_{3/2}}
  = \frac{\langle n^h_{7/2} \rangle + \langle n^h_{5/2} \rangle / [l_v(2l_v-1)]}
         {\langle n^h_{7/2} \rangle + \langle n^h_{5/2} \rangle }
\end{multline}

For the valence states of orbital angular moment $l$,
the expectation value of the angular part of the
spin-orbit interaction is related to the electron occupation numbers
$\langle n_{j_{\pm}}\rangle$
of the total angular momentum levels $j_\pm=l\pm1/2$.
\begin{multline}
\sum_{i \in f} \langle \mathbf{l}_i \cdot \mathbf{s}_i \rangle
=\sum_{j=j_{\pm}} \langle n_j \rangle
  \left[j(j+1)-l(l+1)-\frac{3}{4}\right] \\
=-(l+1)\langle n_{j_{-}}\rangle + l \langle n_{j_{+}}\rangle.
\label{ldots}
\end{multline}

Using the definitions
$\langle n^h \rangle = \langle n^h_{5/2} \rangle + \langle n^h_{7/2} \rangle$
and Eq. (\ref{ldots}), the spin-orbit sum rule for $4d \rightarrow 5f$
transition is given by
\begin{equation}
% B= \frac{\langle n^h_{7/2} \rangle + \langle n^h_{5/2} \rangle / 15}
%         {\langle n^h \rangle }
B  = \frac{3}{5}- \frac{4}{15} \frac{1}{\langle n^h \rangle} \sum_{i \in 5f}
     \langle\mathbf{l}_i \cdot \mathbf{s}_i\rangle.
\end{equation}

While we have made the assumption that the core electrons are
localized, this assumption is not necessary for the valence f
electrons. Therefore under the conditions stated before,
 the sum rule is valid not only in atomic system but also in
solid system, and the spin-orbit strength can be estimated from
the partial occupancy of valence states at a given site,
theoretically. It also shows that the angular part of the
spin-orbit strength is linearly related to the branching ratio in
XAS and EELS \cite{vanderLaan96,thole88b,vanderLaan88,thole88a}.

\begin{figure}[tb]
\includegraphics[angle=270,width=1.0\linewidth]{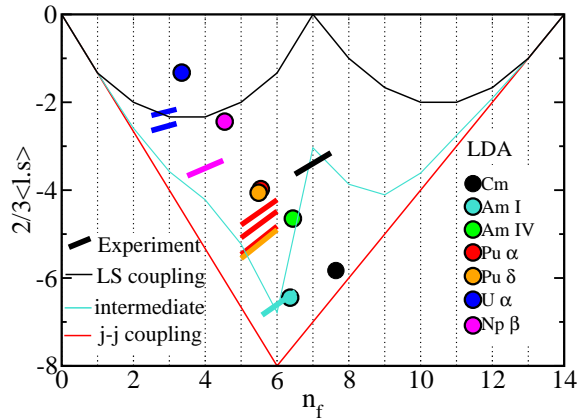}
\caption{
The LDA expectation value of the angular part of the spin-orbit interaction.
The LDA results are denoted by circles for actinide elements as a
function of the $5f$ electron count ($n_f$).
Paramagnetic phase is considered in all cases.
Corresponding experimental data are denoted by the same color of dashes.
The three common angular momentum coupling schemes are shown: $LS$,
$jj$, and intermediate coupling schemes.
}
\label{c_lda}
\end{figure}

\begin{figure*}[tbh]
\includegraphics[angle=270,width=0.8\linewidth]{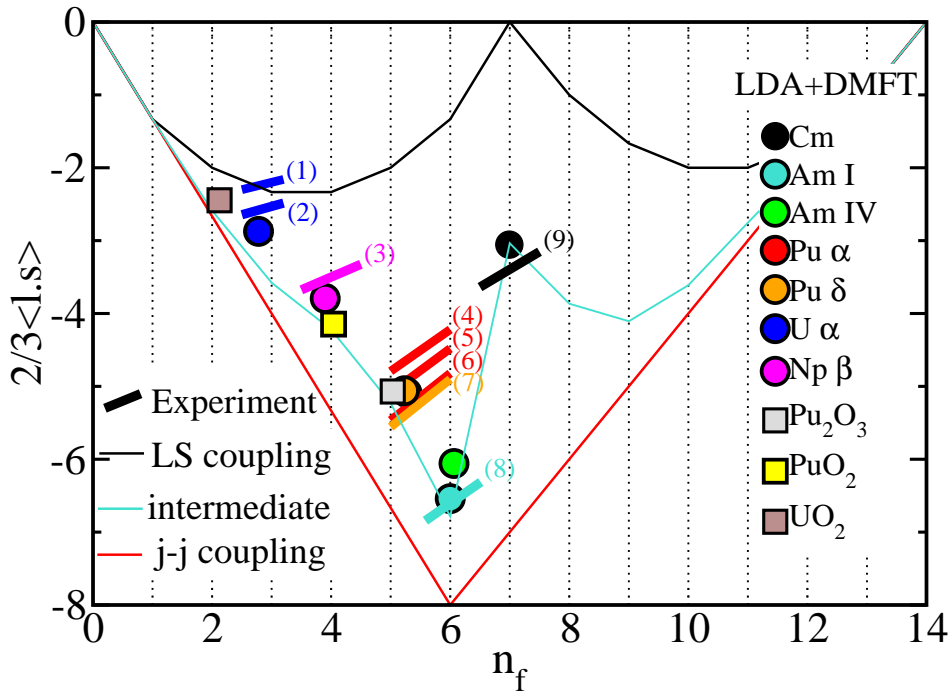}
\caption{
The LDA+DMFT expectation value of the angular part of the spin-orbit 
interaction.
The LDA+DMFT results are denoted by circles and rectangles for actinide
elements and oxides, respectively.
The experimental results obtained from EELS and XAS are denoted by thick
dashes(1-8), because the number of electrons are not defined in the experiments.
(1) $\alpha$-U XAS\cite{Kalkowski87},
(2) $\alpha$-U EELS\cite{vanderLaan04,Moore06},
(3) $\alpha$-Np XAS\cite{Moore07b},
(4) $\alpha$-Pu XAS\cite{vanderLaan04},
(5) $\alpha$-Pu EELS\cite{vanderLaan04},
(6) $\alpha$-Pu EELS\cite{Moore06},
(7) $\delta$-Pu EELS\cite{Moore06},
(8) Am I EELS\cite{Moore07},
(9) Cm EELS\cite{Moore07}.
%$^a$ Ref.~\onlinecite{vanderLaan04},
%$^b$ Ref.~\onlinecite{Kalkowski87},
%$^c$ Ref.~\onlinecite{Moore06},
%$^d$ Ref.~\onlinecite{Moore07},
%$^e$ Ref.~\onlinecite{Moore07b}.
}
\label{c_dmft}
\end{figure*}

%{\bf Implementation Details}

 To obtain the spin-orbit strength and branching ratio,
we calculated the partial occupancies $\langle
n_{j_{\pm}} \rangle$ in solid system by using the LDA and LDA+DMFT
method\cite{rmp_kotliar}. We use a relativistic version of
the linearized muffin-tin orbital (LMTO) method for LDA
calculations\cite{Savrasov96}.
In LDA+DMFT method, the itinerant
$spd$ electrons are treated using LDA, and the strongly
correlated $f$ electrons are considered in DMFT approach, which
maps a lattice problem to a single impurity problem in a self-consistent
electronic bath\cite{rmp_kotliar,rmp_georges}. To
solve the impurity problem, we used the vertex corrected
one-crossing approximation\cite{rmp_kotliar}, 
and the results are
further cross-checked by the continuous time quantum Monte Carlo
method\cite{Haule07}. 
%which is a
%perturbative approach based on the isolated atomic-like $f$-shell.
The Slater integrals $F^k (k=2,4,6)$ and spin-orbit coupling
constants are computed by Cowan' atomic Hartree-Fock (HF) program
with relativistic corrections\cite{Cowan}. We scale the Slater
integrals by 70\% to account for the screening of the solid.
%, which gives better description of branching
%ratio and multiplet structures of photoemission spectra than the
%usual value of 80\% in previous studies\cite{rmp_Moore}. 
We take Coulomb
interaction $U$ = 4.5 eV for actinide elements and 8.0 eV for oxides.
We used an $fcc$ structure for Pu, Am, and Cm with corresponding volume of
each phases, while $\alpha$- and
$\beta$- phases are used for U and Np, respectively.
%This is a reanable approximation because there is little changes in
%the experimental branching ratio
%between different phases of each metal\cite{Moore06,rmp_Moore}.

%{\bf Comparison with LDA}
When the atomic interactions are turned
off, the LDA+DMFT method reduces to the LDA method. Since there has not
been a systematic study of how this method fares vis a vis the
branching ratio, we include some LDA results  in our study.
In Fig.~\ref{c_lda}, we show the branching
ratio of actinide elements obtained by the LDA method and
compare to corresponding experimental data.
We also show results for the three common
angular momentum coupling schemes: $LS$ (Hund's rule is
dominant), $jj$ (spin-orbit interaction is dominant), and
intermediate coupling scheme (obtained from isolated atomic
limit).
The intermediate coupling scheme shows a good agreement with experimental results
of late actinide elements, Am and Cm. In Ref.~\onlinecite{Moore07},
the calculated value of Cm metal overestimates the experimental value.
In our result, it shows better agreement with experiment because
we scaled the Slater integrals by 70\%, which slightly moves the intermediate
coupling scheme toward $jj$ coupling scheme compared to the scaling of 80\%.
The conventional value 80\%
is reasonable approximation for very localized system such as oxides.
However, smaller values are necessary for very itinerant systems.
For example, for transition metal elements, the scaling factor is drastically
decreased for better description of spectroscopy\cite{vanderLaan95}.
As pointed out by Refs.~\onlinecite{vanderLaan04,Tobin05} for $\alpha$-U, and
Ref.~\onlinecite{Moore07b} for $\alpha$-Np,
these metals deviate from the intermediate coupling scheme,
which means these systems are delocalized and the effective spin-orbit
strength decreases.

The LDA results are in overall disagreement with experimental data and
the intermediate coupling scheme.
From U to Pu the spin-orbit strength is much underestimated.
Due to the overestimated band width of the LDA method, the spin-orbit strength
of the actinide element is considerably suppressed compared to the atomic cases.
Only Am is well described by the LDA method due to its special configuration.
There is an optimal spin-orbit stabilization of Am $f^6$ configuration,
which gives clear splitting of occupied $j=5/2$ and uoccupied $j=7/2$ states
in LDA density of states. Our results
show that the LDA description of the branching ratio is not proper for
actinide elements, as shown in Fig.~\ref{c_lda}.

%{\bf DMFT results}
Figure~\ref{c_dmft} shows the branching ratio
obtained by the LDA+DMFT calculation. Our results are in good
agreement with experimental results of all actinide elements. The
branching ratio of late actinides, Am and Cm, are in good
agreement with the intermediate coupling scheme as experimental data.
In well localized system, only single atomic multiplet
configuration is occupied and the system can be well approximated by an
atomic problem, which is the intermediate coupling scheme.
On the other hand, as the system is delocalized the $f$
electrons become fluctuating between various atomic
configurations and exchanging electrons with the surrounding
medium. As a result, the spin-orbit strength is suppressed and
the branching ratio moves from the intermediate scheme towards 
the $LS$ coupling scheme.
Note that the value for U and Np in
Fig.~\ref{c_dmft} is further away from the intermediate coupling scheme
than late actinides. Also
note that this is not a competition between the spin-orbit
iteraction and the Hund's rule coupling. This is the competition
between the spin-orbit interaction and the delocalization. If the
system is fully delocalized, the spin-orbit strength goes
to zero and the branching ratio is just statistical value of
3/5. According to our plot, $\delta$-Pu also deviates from the
intermediate coupling scheme, hence this system is partly itinerant.
This deviation becomes larger for earlier actinides such as Np and U.

Our results for U and Np slightly overestimate the spin-orbit strength
measured in experiments.
Here, we neglected the interference effect between different site in the
description of core-valence transition as expalined in Eq.~(4).
In the impurity problem, we use the one crossing approximation method,
which might overestimate
the localization of $f$ electrons because it is based on the local atomic
description. Also, the on-site Coulomb interaction $U$ should be smaller
than 4.5 eV for rather itinerant system, $\alpha$-U and $\alpha$-Np.
By improving those effects, LDA+DMFT can
give more delocalized branching ratio behavior.

We also investigate the change of spin-orbit strength
under pressure, which are not available in atomic calculation.
The volume of $\alpha$-Pu and Am IV is highly suppressed by
25\% and 40\% compared to the volume of $\delta$-Pu and Am I, respectively.
%For the description of $\alpha$-Pu and Am IV, whose volumes are highly
%suppressed by 25\% and 40\% from the volume of the ambiant pressure,
%we simply decreases the volume with $fcc$ structural phase.
The difference of $n_f$ and the branching ratio between $\alpha$-Pu and
$\delta$-Pu is almost negligible and hard to detect, similar to EELS
experiments\cite{Moore06,rmp_Moore}. On the other hand,
there is a noticable change between Am I and Am IV.
There is a transition from
optimal spin-orbit stabilization for $f^6$ to optimal exchange
interaction stabilization for $f^7$ configuration\cite{Moore07}.
This transition induces a significant change in branching ratio between 
Am I and Am IV only with slight change of $n_f$.
%The angular momentum coupling plays a decisive role in
%the formation of magnetic moment in Cm metal.

\begin{figure}[tb]
\includegraphics[angle=270,width=1.0\linewidth]{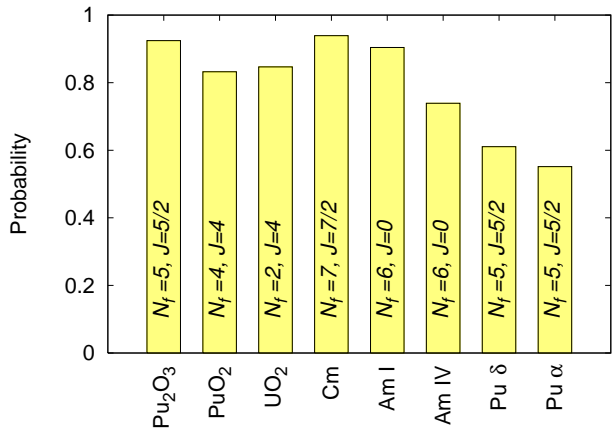}
\caption{
Probability of the most occupied atomic ground state multiplet.
The height of the peak corresponds to the fraction of the time 
the $f$ electrons of the solid spends in the given atomic multiplet,
denoted by the $5f$ electron count $N_f$ and the total spin $J$ of the atom.
%{\bf We didn't put the alpha-U and beta-Np}
}
\label{nf}
\end{figure}

The branching ratio shows systematic deviation from atomic calculation
across the actinide series.
Figure~\ref{nf} shows the probability of the most occupied ground state
multiplet for each elements.
The branching ratio of Cm and Am are very close to the atomic case,
because the probability of the $f^7$ ($f^6$) ground state is 95\% (90\%)
in valence histogram\cite{Shim},
which reflects almost atomic ground states.
As the atomic number is decreased, the system becomes delocalized,
and the spin-orbit strength is decreased.
In early actinides, the $5f$ electrons are delocalized
and they are distributed over various atomic configurations and
not only in the ground state atomic multiplet.
Note that $\delta$-Pu already shows deviation from localized states,
as discussed in the suppressed spin-orbit strength of $\delta$-Pu.
Under pressure, the $5f$ electrons are delocalized, and the
probability of atomic ground state multiplet is decreased
as shown in Am IV and $\alpha$-Pu.

%The $n_f$ in LDA+DMFT caculations shows small
%deviation from integer number expected from trivalent ion.
%Am and Cm, which are localized system, shows integer number of $n_f$.
%However, the others shows additional $f$ electrons over the atomic occupancy.
%Especially, the pressurized phase, $\alpha$-Pu and Am(IV) shows additional
%increase of $n_f$ by pressure.

%This indicates that pressure in general enhances the number of $f$ electrons.
%However, we should be careful for the exact value of $n_f$,
%because it is very sensitive to the double counting scheme and
%the Coulomb interaction parameter $U$.
%For example, if we choose $U$ = 4.0 eV for $\delta$-Pu,
%the $n_f$ and $B$ changes to 5.28.

Finally  we also study actinide oxide systems Pu$_2$O$_3$,
PuO$_2$, and UO$_2$.  These materials are very important because
they are used in nuclear fuels for energy generation.
LDA+DMFT predicts that  these materials are   localized. As shown
in Fig.~\ref{c_dmft}  their branching ratios are  consistent with the
intermediate coupling scheme. The calculated $n_f$ indicates that trivalent
metal ion in Pu$_2$O$_3$ and tetravalent metal ion in PuO$_2$ and UO$_2$,
which shows that these systems can be described by ionic system
rather than metallic or covalent system.
Recent experiments on the $4d\rightarrow 5f$ transition in EELS
show that the branching ratio of UO$_2$  and PuO$_2$ are similar
to that of $\alpha$-U and $\delta$-Pu, respectively and it has
been  suggested that these actinide dioxides have covalent
metal-oxide bonding, and the number of $f$ electrons can become
non-integer number\cite{Moore06}.
However, unpublished EELS results of the $5d\rightarrow 5f$ transition
in UO$_2$  and PuO$_2$ suggest the bonding is indeed ionic with
a near integer change in $5f$ occupancy\cite{Moore}.
So, further experimental work is required to investigate
why the $5d$ and $4d$ spectra show ionic and covalent bonding nature,
respectively.
%
%Theses experiment is in disagreement with our LDA+DMFT results showing
%clear ionic feature.

%There is several possibility for the disagreement with our results and
%experiment.
%We assume that the localized scheme, where the interference
%between different site is negligible. However, we can expect that
%PuO2 is more localized than Pu, so we can exclude this possiblity.
%Also we use an approximation that there is no core-valence interaction.
%This is expected to be small in actinide and it cannot be a reason.
%Reconciling this difference, may require further experimental work
%to rule out the  possiblity of the presence of  Pu$_2$O$_3$, which as
%shown in our results, Pu$_2$O$_3$ give similar result to the
%experimental results of $\delta$-Pu.

Acknowledgment: We acknowledge useful discussions with K. T. Moore.


\begin{thebibliography}{99}


\bibitem{Arko00} A. J. Arko {\it et al.},
        Phys. Rev. B {\bf 62}, 1773 (2000).
\bibitem{Tobin03} J. G. Tobin {\it et al.},
        Phys. Rev. B {\bf 68}, 155109 (2003).
\bibitem{Gouder05} T. Gouder {\it et al.},
        Phys. Rev. B {\bf 71}, 165101 (2005).
\bibitem{Shim} J. H. Shim, K. Haule, and G. Kotliar,
        Nature {\bf 446}, 513 (2007).
\bibitem{Shick07} A. Shick {\it et al.},
        Europhys. Lett. {\bf 77}, 17003 (2007).
\bibitem{Zhu07} J. X. Zhu {\it et al.},
        Phys. Rev. B {\bf 76}, 245118 (2007).
\bibitem{ctqmc} C. A. Marianetti, K. Haule, G. Kotliar, and M. J. Fluss,
        Phys. Rev. Lett. {\bf 101}, 056403 (2008).
\bibitem{vanderLaan04} G. van der Laan {\it et al.},
        Phys. Rev. Lett. {\bf 93}, 097401 (2004).
\bibitem{Tobin05} J. G. Tobin  {\it et al.},
        Phys. Rev. B {\bf 72}, 085109 (2005).
\bibitem{Moore07b} K. T. Moore, G. van der Laan, M. A. Wall, A.J. Schwartz, and R. G. Haire.
        Phys. Rev. B {\bf 76}, 073105 (2007).
\bibitem{Moore07} K. T. Moore  {\it et al.}
        Phys. Rev. Lett. {\bf 98}, 236402 (2007).
\bibitem{vanderLaan96} G. van der Laan and B. T. Thole,
        Phys. Rev. B {\bf 53}, 14458 (1996).
\bibitem{Moore06} K. T. Moore  {\it et al.}
        Phys. Rev. B {\bf 73}, 033109 (2006).
\bibitem{Butterfield08} M. T. Butterfield, K. T. Moore, G. van der Laan, M. A. Wall, and R. G. Haire,
        Phys. Rev. B {\bf 77}, 113109 (2008).
\bibitem{rmp_Moore} K. T. Moore  and G. vand der Laan,
	arXiv:0807.0416 (2008)
\bibitem{Moore04} K. T. Moore  {\it et al.}
        Philos. Mag. {\bf 84}, 1039 (2004).
\bibitem{thole88b} B. T. Thole and G. van der Laan,
        Phys. Rev. B {\bf 38}, 1358 (1988).
\bibitem{vanderLaan88} G. van der Laan and B. T. Thole,
        Phys. Rev. Lett. {\bf 60}, 1977 (1988).
\bibitem{thole88a} B. T. Thole and G. van der Laan,
        Phys. Rev. A {\bf 38}, 1943 (1988).
\bibitem{vanderLaan98} G. van der Laan,
        Phys. Rev. B {\bf 57}, 112 (1998).
\bibitem{rmp_kotliar} G. Kotliar {\it et al.},
        Rev. Mod. Phys. {\bf 78}, 865 (2006).
\bibitem{Savrasov96} S. Y. Savrasov,
        Phys. Rev. B {\bf 54}, 16470 (1996).
\bibitem{rmp_georges} A. Georges, G. Kotliar, W. Krauth, and M. J. Rozenberg,
        Rev. Mod. Phys. {\bf 68}, 13 (1996).
\bibitem{Haule07} K. Haule,
	Phys. Rev. B {\bf 75}, 155113 (2007).
\bibitem{Cowan} R. D. Cowan,
        {\it The Theory of Atomic Structure and Spectra}
        (Univ. California Press, Berkeley, 1981).
\bibitem{Kalkowski87} G. Kalkowski, G. Kaindl, W. D. Brewer, and W. Krone,
        Phys. Rev. B {\bf 35}, 2667 (1987).
\bibitem{vanderLaan95} G. van der Laan,
        Phys. Rev. B {\bf 51}, 240 (1995).
%\bibitem{Baer80} Baer and Lang
%        Phys. Rev. B {\bf 21}, 2060 (1980).
\bibitem{Moore} K. T. Moore, (personal communication).

\end{thebibliography}
\end{document}